\documentclass[a4paper,11pt]{article}
\pdfoutput=1 

\usepackage{jinstpub} 
\usepackage{textcomp}  

\title{\boldmath The Cryogenics and Xenon handling system for the PandaX-4T Experiment}

\author[a,b,1]{Li Zhao,\note{Corresponding author.}}
\author[b,c,1]{Xiangyi Cui,}
\author[a]{Wenbo Ma,}
\author[d]{Yingjie Fan,}
\author[a]{Karl Giboni,}
\author[a,b]{Tao Zhang,}
\author[a,b,c]{Jianglai Liu,}
\author[a,b,c]{Xiangdong Ji}

\affiliation[a]{INPAC, School of Physics and Astronomy, Shanghai Jiao Tong University,
             \\Shanghai Key Laboratory for Particle Physics and Cosmology, Shanghai 200240, China}
\affiliation[b]{Shanghai Jiao Tong University Sichuan Research Institute, Chengdu 610213, China}
\affiliation[c]{Tsung-Dao Lee institute, Shanghai Jiao Tong University, Shanghai, 200240, China}
\affiliation[d]{School of Physics, Nankai University, Tianjin 300071, China}

\emailAdd{zhaoli78@sjtu.edu.cn, hongloumeng@sjtu.edu.cn}

\abstract{PandaX-4T, the new generation of the PandaX detector, is a xenon dual-phase detector with about six tons of liquid xenon as target for dark matter search. A cryogenics and xenon handling system is designed to liquify and purify this large detector. In this paper, the results on the commission of the cryogenics and gas handling system are reported. The maximum cooling power of $\sim$ 580 W at 178 K with three cooperating coldheads has been achieved. The filling rate with an assisted liquid nitrogen cooling can reach $\sim$ 700 kg/day. The average rate of recuperation by liquid nitrogen is around 440 kg/day. The maximum total purification speed of two circulation loops is up to $\sim$ 155 slpm. Each loop is using a large heat exchanger with a measured heat exchange efficiency of $\sim 97.5\pm0.5\%$.}


\begin{document}
\maketitle

\section{Introduction}
The worldwide hunt for dark matter in the form of Weekly Interacting Massive Particles(WIMPs) is accelerated by the remarkable progress of liquid xenon time projection chambers. In the past a few years, LUX\cite{LUX}, XENON\cite{XENON1T} and PandaX\cite{PandaX-I-2014,PandaX-I-2015,PandaX-II-2016-comi,PandaX-II-2016,PandaX-II-2017,PandaX-rev-2018,PandaX-II-2020} have produced stringent limits on the WIMP-nucleon intersections. XENON1T experiment is being upgraded to XENONnT\cite{XENONnT} with $\sim$ 8
tons of xenon; LUX and ZEPLIN\cite{ZEPLIN-III} are being upgraded to LZ\cite{LZ} with $\sim$ 10 tons of xenon. The next phase of PandaX dark matter program, PandaX-4T\cite{PandaX-4T-future,PandaX-4T-sim}, will use about 6 tons of xenon at China JinPing underground Laboratory phase II (CJPL-II)\cite{CJPL-II}.

Holding and purifying a large amount of liquid xenon in PandaX-4T is a big challenge. Operating such a massive liquid xenon detector has not been reported in the literature to-date. For PandaX-II with $\sim$1100 kg of liquid xenon, the gas circulation rate was limited to around 30 slpm\cite{Cry-Pandax-II} by its cryogenics (available cooling power: $\sim$180 W). The maximum purification flow of XENON1T ($\sim$ 3 tons of xenon) was 114 slpm\cite{Cry-XENON1T}. The xenon mass turn-over time of LUX ($\sim$ 370 kg of xenon) was about 40 hours\cite{Tri-LUX}, the corresponding flow rate was around 28 slpm. For PandaX-4T, we plan to increase the recirculation speed to be around 200 slpm, which corresponds to $\sim$ 1600 kg/day. Therefore, a new cryogenic system was designed, constructed and tested for it. It mainly consisted of a Cooling Bus\cite{Cry-Pandax-II}, heat exchanger (HE) modules, circulation pumps, hot purifiers and other supporting system. All the components were chosen to handle high flow, including pipes, valves, heat exchangers (HEs), pumps and purifiers. In this paper, the HE efficiency and flow characteristics of the purifier have been measured. The performance of the Cooling Bus is reported in details.

\section{Experimental setup}
\subsection{The supporting infrastructure}
\begin{figure}[bhtp]
\centering
\includegraphics[width=1.0\textwidth]{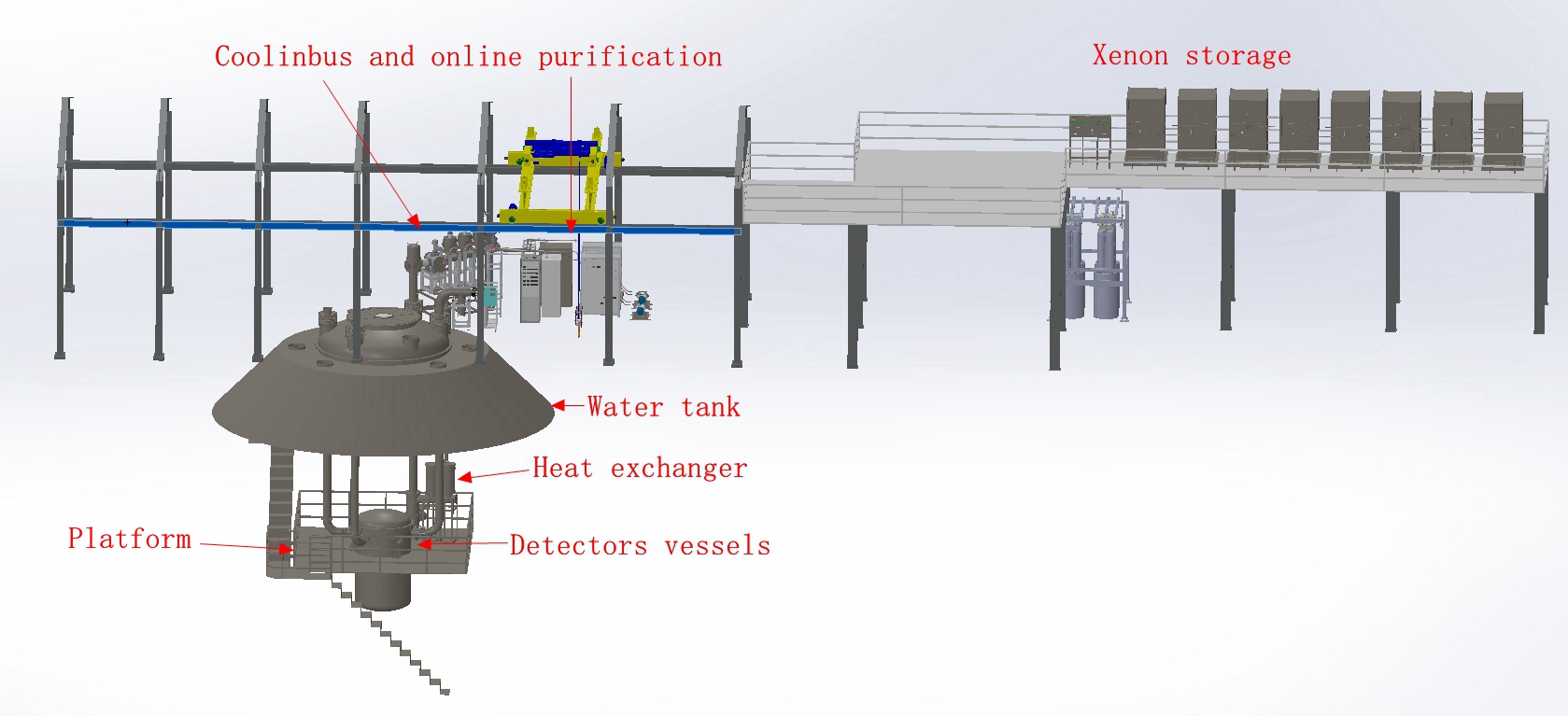}
\caption{Layout of the PandaX-4T detector vessels and its supporting systems at CJPL-II.}
\label{fig:layout-pandax4t}
\end{figure}

Figure~\ref{fig:layout-pandax4t} depicts the layout of the detector vessels and its
supporting infrastructure. Key elements include the detector vessels located inside an ultra pure water tank, the Cooling Bus, xenon online purification system, the underwater HEs and xenon storage.
In order to suppress the background of the detector, the water tank (diameter: $\sim$10 m; height: $\sim$13.5 m) is built in a big dry pool pit as a water shielding. It can hold $\sim$900 tons of water. The detector vessels (outer and inner) are placed at the center of the water tank, about 7 m below the lab floor. The cryostat vessel (surface:$\sim$ 13 $m^{2}$) with 10 layers of aluminized Mylar foil is supported with 3 Torlon feet inside the outer vessel. The detector vessels are made of the radiopure stainless steel\cite{pure-steel}.

\subsection{Xenon storage and emergency recovery system}
\begin{figure}[bhtp]
\centering
\includegraphics[width=0.6\textwidth]{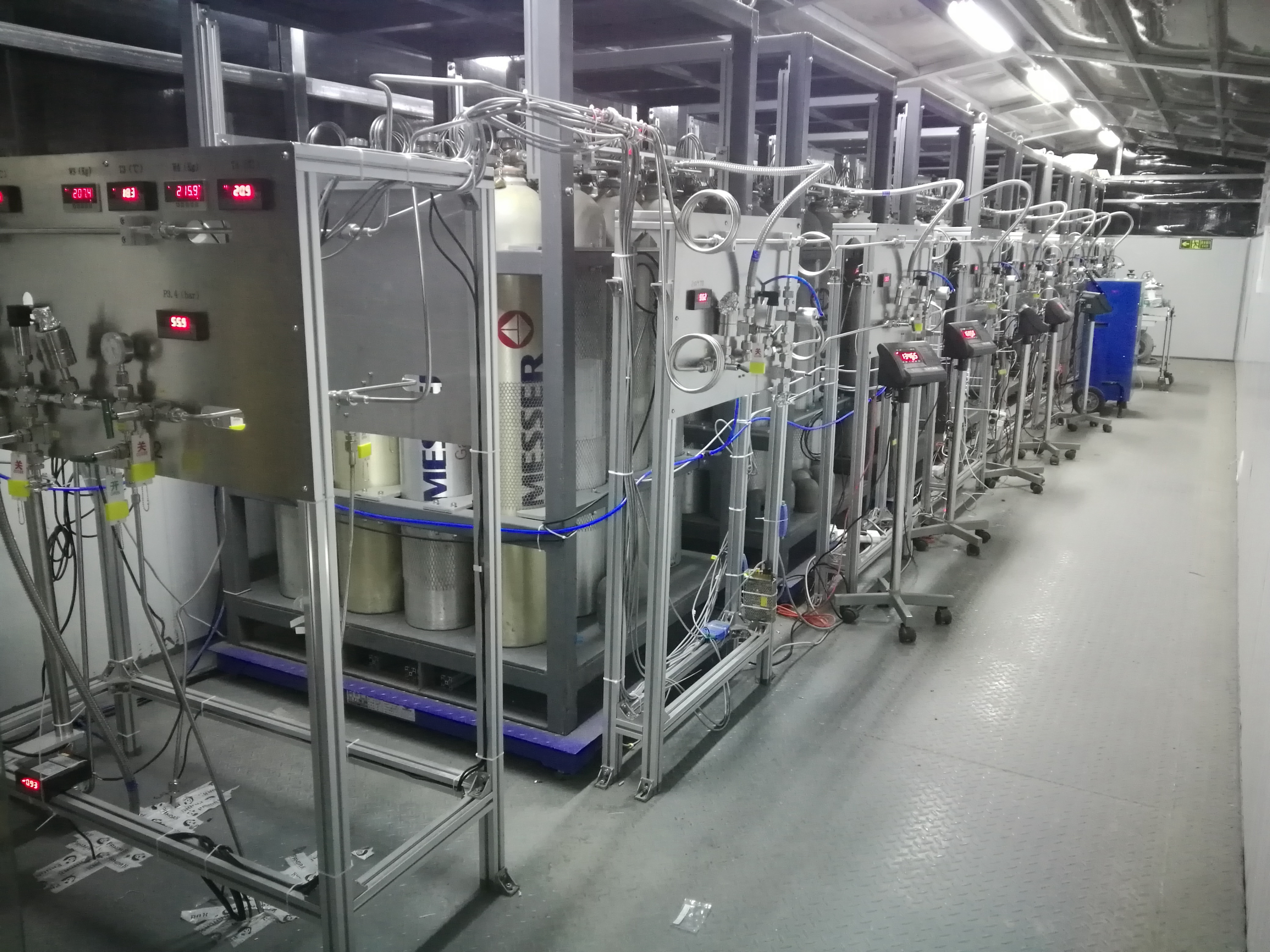}
\includegraphics[width=0.315\textwidth]{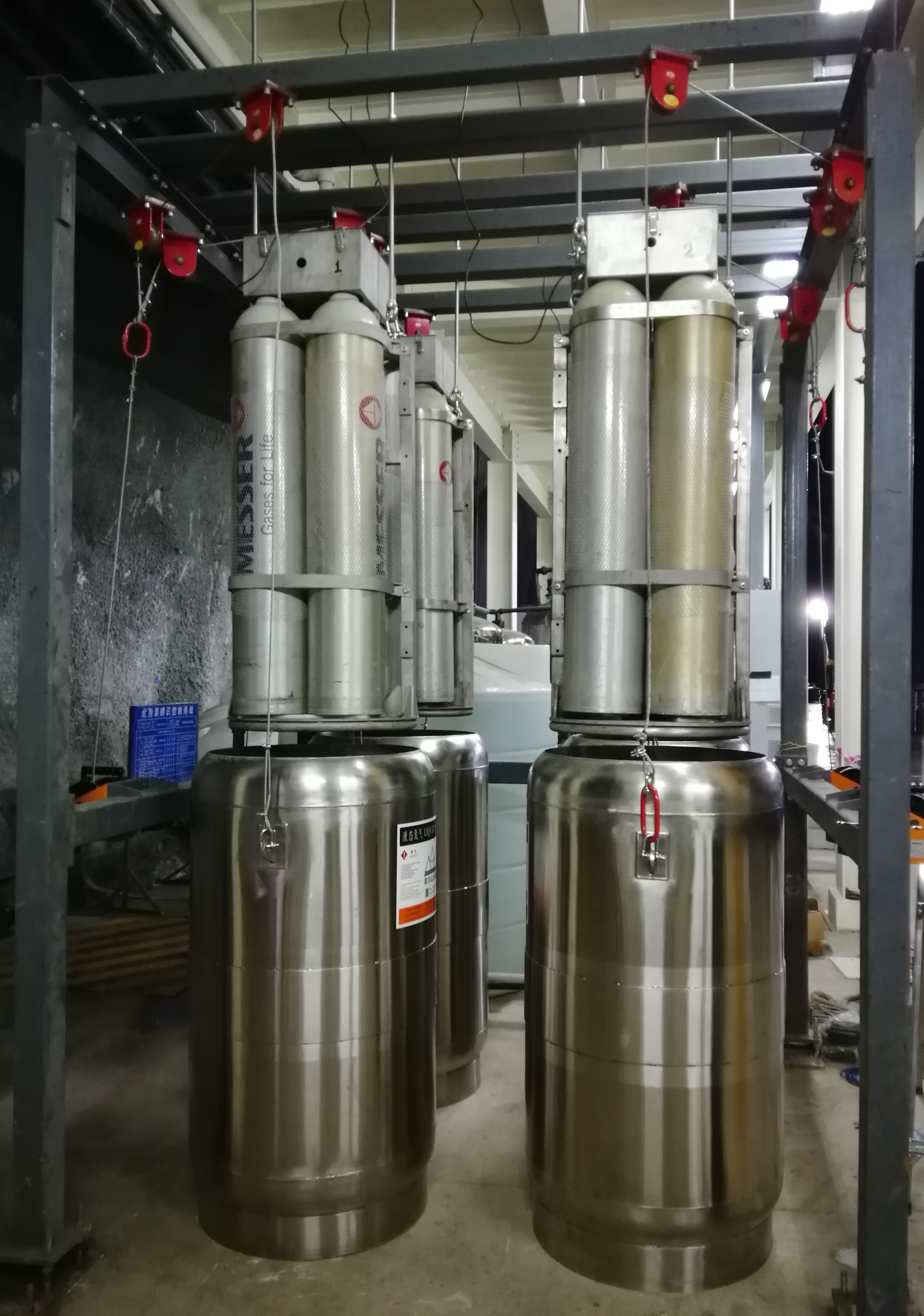}
\caption{Xenon room temperature storage (Left) and LN2 emergency recovery system (Right) at CJPL-II.}
\label{fig:RoomT-storage-LN2-recov}
\end{figure}

Xenon room temperature storage system (Figure~\ref{fig:layout-pandax4t}) is about 20 m far away from the detector running region. The xenon storage mainly consists of 128 gas cylinders (40-liter) with D200-Brass cylinder valves (ROTAREX, Luxembourg), manifolds and packs. Each pack has 16 cylinders connected with each other. A leak check and vacuum pumping has to be performed for each pack before filling xenon. The packs are stored in a gas room on the second floor of a steel platform (Figure~\ref{fig:RoomT-storage-LN2-recov}). Each pack has a control panel, through which all packs are connected with each other. The panel has a pressure sensor,
a rupture disc, a high pressure bellow, some manual valves and
a high pressure pneumatic valve, which makes it easy to disconnect and
connect one pack from the whole gas system. The pneumatic valve is for
remote operation. Manual valves are for both normal and emergency
case. Moreover, their ambient temperature and gas pressure are monitored and
recorded by slow monitor system.

The emergency recovery system by liquid nitrogen (LN2) is located on the first floor of the steel platform. It would be turned on manually if there is a need for xenon recuperation. Figure~\ref{fig:RoomT-storage-LN2-recov}
shows its structure. Four groups of 4 cylinders are housed in a pre-engineered
pack that can be handled with a crane. Each cylinder (40-liter) is sealed with a steel cylinder valve D304 (ROTAREX, Luxembourg), which is marketed to the semiconductor industry for ultra-high-purity gas application and is different from the valve used for storage cylinder. The cylinder outlet valve connection is DISS 718, VCR-like fitting that makes an all-metal seal. The cylinders in the pack are
connected with a stainless steel manifold that mates to the DISS fittings. There is a control panel for each pack, like that of the storage system. Their key parameters are also monitored and recorded. Frozen xenon in these 4 packs can be pushed into the storage cylinders by warming them up, which is the way to recuperate massive xenon. Therefore, 6 tons of xenon can be recuperated continuously, each time by keeping two packs in LN2, and the other two in room temperature alternatively.

\subsection{The Cooling Bus with multiple coldheads}
Due to central location of PandaX-4T detector in the water tank, cryogenics has to be connected to the detector vessels with long pipes through the water shielding. The Cooling Bus and circulation devices are located on the lab floor for easy maintenance.

A 100 mm diameter gas pipe with a vacuum insulated jacket is chosen to
transport the gas between the inner vessel and the Cooling Bus, and it is also a
vacuum pumping line for the inner chamber including the detector before filling
xenon into it. The xenon, which is liquified by coldheads, can run back in a
separate concentric 16 mm diameter pipe. The liquid xenon in this pipe is
driven by gravity, and the liquid pipe is therefore mounted at a $5^{\circ}$
angle in the Cooling Bus, then goes out the gas pipe in the join vertical section and
down into the inner vessel separately. They are enclosed in a 250 mm
diameter vacuum pipe with aluminized Mylar foils to minimize
heat leaks from the outside walls. For all the devices connected to the
detector vessels, the 'L' shape pipes (Figure~\ref{fig:layout-pandax4t}) in the water shielding are chosen.

\begin{figure}[bhtp]
\centering
\includegraphics[width=0.7\textwidth]{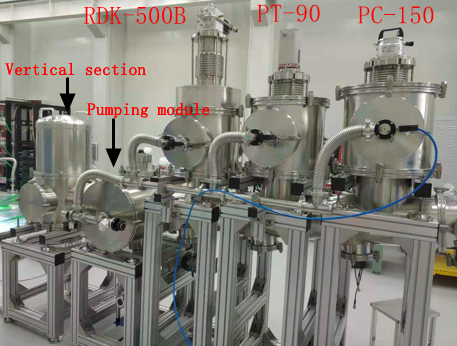}
\caption{The Cooling Bus is assembled in the lab.}
\label{fig:coolinbus-real}
\end{figure}

Figure~\ref{fig:coolinbus-real} shows the Cooling
Bus with 3 coldheads (a single stage Gifford McMahon (GM) RDK-500B\cite{RDK-500B} (Sumitomo Heavy Industries, Japan), a Pulse Tube Refrigerator(PTR) PT-90\cite{PT-90} (Cryomech, USA), a PTR PC-150\cite{PC-150} (JEC, Japan)). All cooling towers on the Cooling Bus have the same length and angle,
and can be equipped with different coldheads. The cooling power of coldheads
is delivered to the xenon in the gas pipe through a cold finger made of Oxygen-Free High-Conductivity (OFHC ) copper installed at the top of the xenon gas chamber. Fins machined in the cold finger
increase the surface of liquifying xenon gas. The
drops of liquid xenon condensed on the cold finger fall into a funnel which
guides them into the central liquid pipe. The cold finger is sealed with an
indium wire to a 150 mm diameter cylinder vessel, and the inner
chamber of this vessel is connected to the xenon gas pipe with a 50 mm
diameter bellow. The bottom flange of the vessel is extended to the outer
chamber of the cooling tower with 2 mm thickness stainless steel cylinder.
Thus, the vacuum chamber of coldheads is separated from that of the Cooling Bus. Therefore, it is convenient to service or replace without opening
the inner xenon vessel and breaking the outer vacuum chamber of the detector. An additional cooling loop contains a stainless steel tube wound into a coil, which goes through inner and outer
chamber. LN2 flowing through the coil condenses xenon gas and the liquid
will drop into a funnel similar to the coldhead. Such coil is installed in the
cooling tower of RDK-500B. In short, all the coolers of the Cooling Bus do not come in contact with
xenon directly, neither do their auxiliary parts (sensors, heaters and cables).

In the 500B cooling tower, RDK-500B coldhead is equipped with an OFHC copper
adapter and mounted on the copper finger. Pt100 sensors are put in the holes of the copper finger. Cartridge heaters\cite{Lakeshore-pro} are inserted into suitable holes of the adapter. Inner structure of PT-90 cooling tower is similar. However, that of PC-150 tower is a little different. It has a cup-shaped electrical heater (provided by
the company) between the coldhead and the copper adapter. All the Pt100 sensors of the Cooling Bus are read by two temperature controllers (Lakeshore 350, USA) \cite{Lakeshore-pro}, which control the power to the heaters and keep the cold finger stable at
a set temperature.

The long 'L' shape pipe above, just outside the water tank, is connected to the Cooling Bus by a vertical section, which allows rotation around the center of the pipe, so the Cooling Bus can be installed in nearly any direction. The next section is a vacuum pumping module with a 400 mm diameter outer chamber, in which the inner pipes can be connected easily. Its bottom is connected to a turbopump (KYKY FF-200/1200, China) via a 200 mm diameter gate pneumatic valve. The turbopump is backed by a dry scroll pump (Agilent IDP-15, USA). This set of pumps is for the insulation and cryostat vacuum. On the top, a turbopump (Agilent TV304FS, USA) is mounted for inner chamber via a all metal manual angle valve. The angle valve would be closed before filling the detector. Its forepump is IDP-15 too. Inner pressure sensor, rupture discs, and outer gauges are set up for the Cooling Bus.

Considering a lot of aluminized Mylar foil and the inner tubes in 'L' shape vacuum pipe, the second pumping station with a LN2 sorption pump is added and connected to the outer vessel directly with a 250 mm diameter empty pipe to maintain good outer
vacuum. The pumping station is also equipped with the pumps, a gate
valve and gauges, which are the same to that of the Cooling Bus's outer chamber. The LN2 sorption pump is for emergency cases, such as power-off or malfunctioning of pumps.

\subsection{Online purification system}
\label{sec:online}
\begin{figure}[bhtp] \centering
\includegraphics[width=0.8\textwidth]{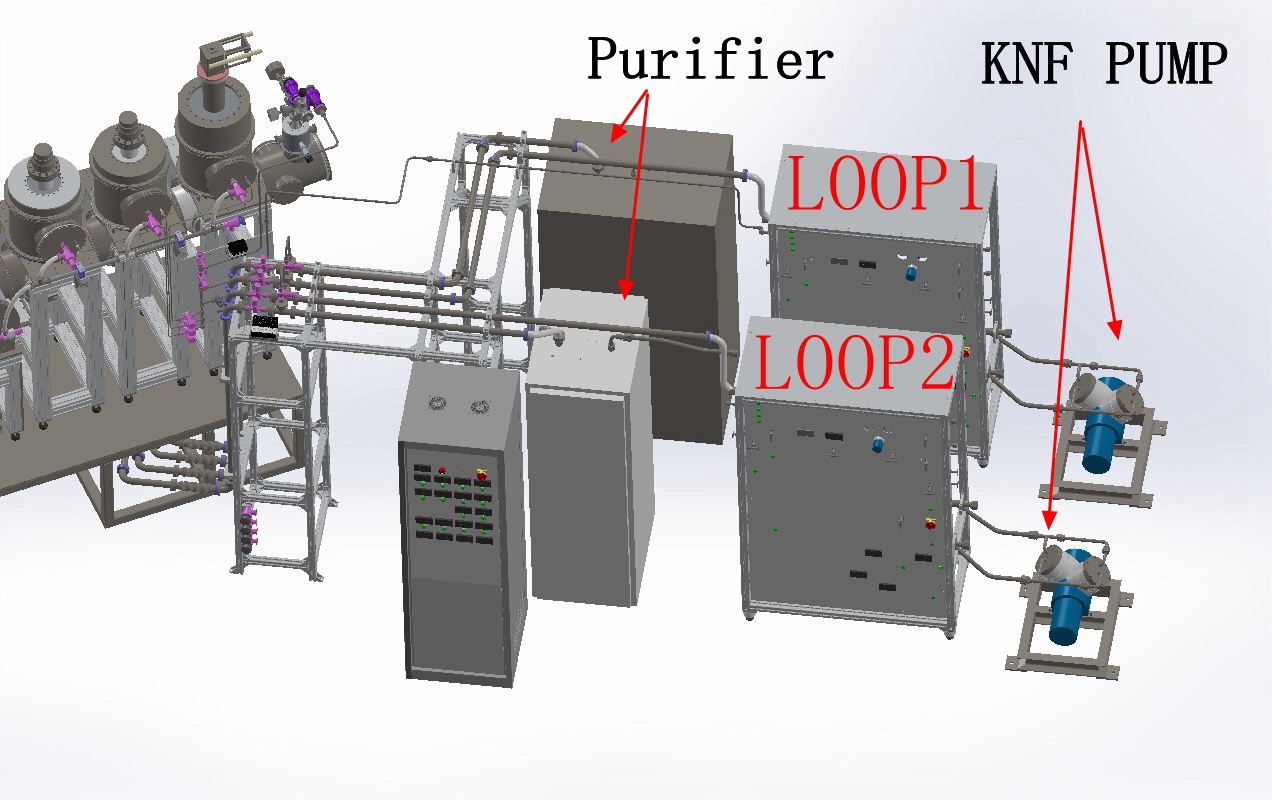}
\caption{The schematic design of xenon gas purification system with two parallel loops.}
\label{fig:loop-purifying}
\end{figure}

Normally, the impurities of liquid xenon in a large detector mainly come from material outgassing. Although the vessel and detector components can be pumped and baked-out to suppress the outgassing, residual outgassing remains. Therefore, an online purification system with high speed is specially designed for PandaX-4T. Figure~\ref{fig:loop-purifying} shows the schematic design. There are two parallel loops (LOOP1 and LOOP2) to run independently to minimize interruption of the detector. Each loop mainly includes a HE in water (Figure~\ref{fig:layout-pandax4t}), a circulation pump (KNF pump), and a hot purifier.

The heat exchanger is a standard brazed plate HE,
model K205-50W-NB93\cite{Kaori-HE-205} (KAORI, China), which consists of 50 plates (size: 528 mm*246 mm*131.5 mm, with a volume of about 11 $l$ and total heat transfer area of $\sim$ 5
$m^{2}$) and withstands $100\sim300$ slpm. The circulation pump is a large
capacity double-headed diaphragm pump, model PM28943-N1400.12\cite{KNF-1400} (KNF, Germany), which has two fold safety by the combination of a working diaphragm and an additional
safety diaphragm. It is normally capable of flowing $\sim 250$ slpm of air
at atmospheric pressure on the input, and its maximum operating pressure on the
output is 3.0 barg. Two pressure gauges are mounted at the input and output of the KNF pump for each loop. The flow meter between the KNF pump and the purifier is a FM5012\cite{FM-5000} (Siargo, USA), calibrated up to 300 slpm of xenon gas. The purifier of LOOP1 is a model PS5-MGT50-R-909\cite{PS5-Getter} (Entegris, Singapore), rated for purification of rare gas at flow of 100 slpm. The purifier of LOOP2 is a model 9N300-R\cite{9N300-Getter} (Simpure, China), which can stand 100 slpm of xenon gas flow. The main gaslines of the circulation system are made of 35 mm diameter stainless steel pipes with CF35 flanges, to allow the rapid flow through the loops. Moreover, for self purification of loops, a pneumatic valve is set between the outlet and the inlet of each loop panel.

Most xenon is purified by continuous gas circulation through the
hot purifier, taking liquid xenon from the overflow chamber into the HE and letting the purified xenon gas cool and liquify inside the HE on its way back into the detector.
The overflow chamber is located at bottom of the inner vessel. Due to high density of liquid xenon and limited capability of KNF pump, the HE models with an insulated vacuum chamber are positioned on the the platform in the water shielding close to the overflow chamber, which is helpful to take liquid xenon into the HE by pressure difference. This is a major difference from that of PandaX-II\cite{Cry-Pandax-II}. Considering impurities (H2, N2, O2, CO, CH4) with lower boiling point, which can be easily enriched in upper gas xenon ($\sim$ 9 m above the liquid level), a fraction of xenon gas ($\sim$ 5 slpm) near the copper finger in the Cooling Bus is guided into LOOP1 for purification.

\section{Commissioning results and discussion}
\subsection{Performance of the Cooling Bus}

\begin{figure}[bhtp] \centering
\includegraphics[width=0.8\textwidth]{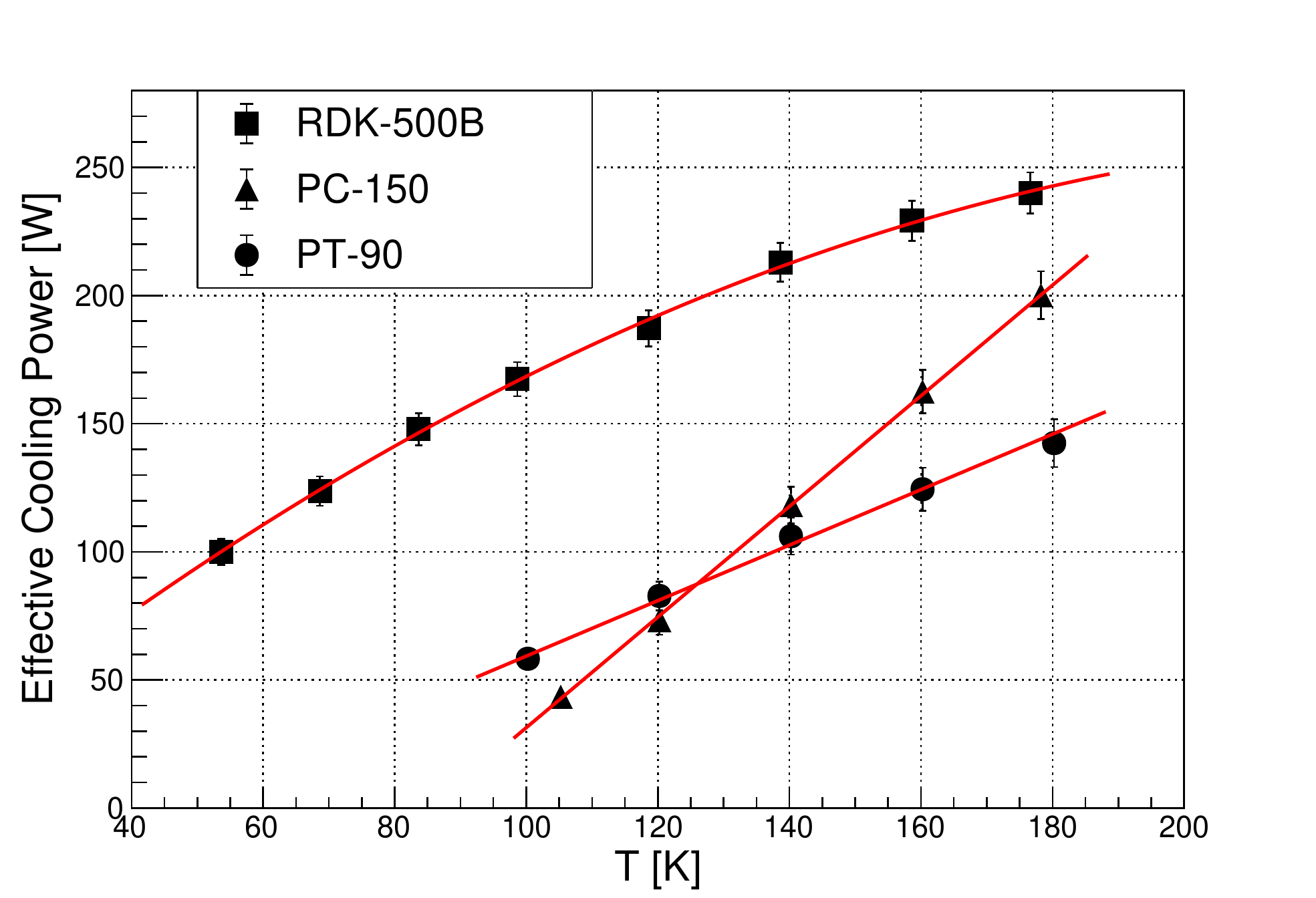}
\caption{The effective cooling power of coldheads on the Cooling Bus at different temperature of the copper
finger.} \label{fig:coolingpower-bus}
\end{figure}

The effective cooling power of coldheads on the Cooling Bus has been tested as
the inner and outer chamber in a good vacuum(<$1\times10^{-3}$ Pa). The Lakeshore 350 Proportional, Integral and Derivative(PID) controller reads the temperature at the cold finger and controls the electric power to the heaters. However, the maximum power of heating of the controller is not sufficient to counteract the cooling power at all times. So, the secondary control loop of 350, providing a $0\sim10$ V DC signal, controls a 500 W DC power supply of the heaters, the temperature of the cold finger can be stable at the set value in an hour. The values of PID are set according to the experience of PandaX-II. Performance of one coldhead is quite stable as other coldheads are running or stopped in the test. The test results are shown in Figure~\ref{fig:coolingpower-bus}. The effective cooling power of GM RDK-500B coldhead is $\sim$ 240 W at 178 K, that of PTR PC-150 and PT-90 is $\sim$ 200 W and $\sim$ 140 W respectively. The cooling power of the PC-150 is higher than that of the PT-90 at higher temperature points ($>$130 K). The cooling power of the PC-150 at 178 K is a little higher than that ($\sim$ 180W) of PandaX-II\cite{Cry-Pandax-II} because of different Helium compressor (4.3 kW
water-cooled Cryomech CP2850). The total cooling power of the Cooling Bus with these three coldheads is $\sim$580 W at 178 K.

\begin{figure}[bhtp] \centering
\includegraphics[width=0.8\textwidth]{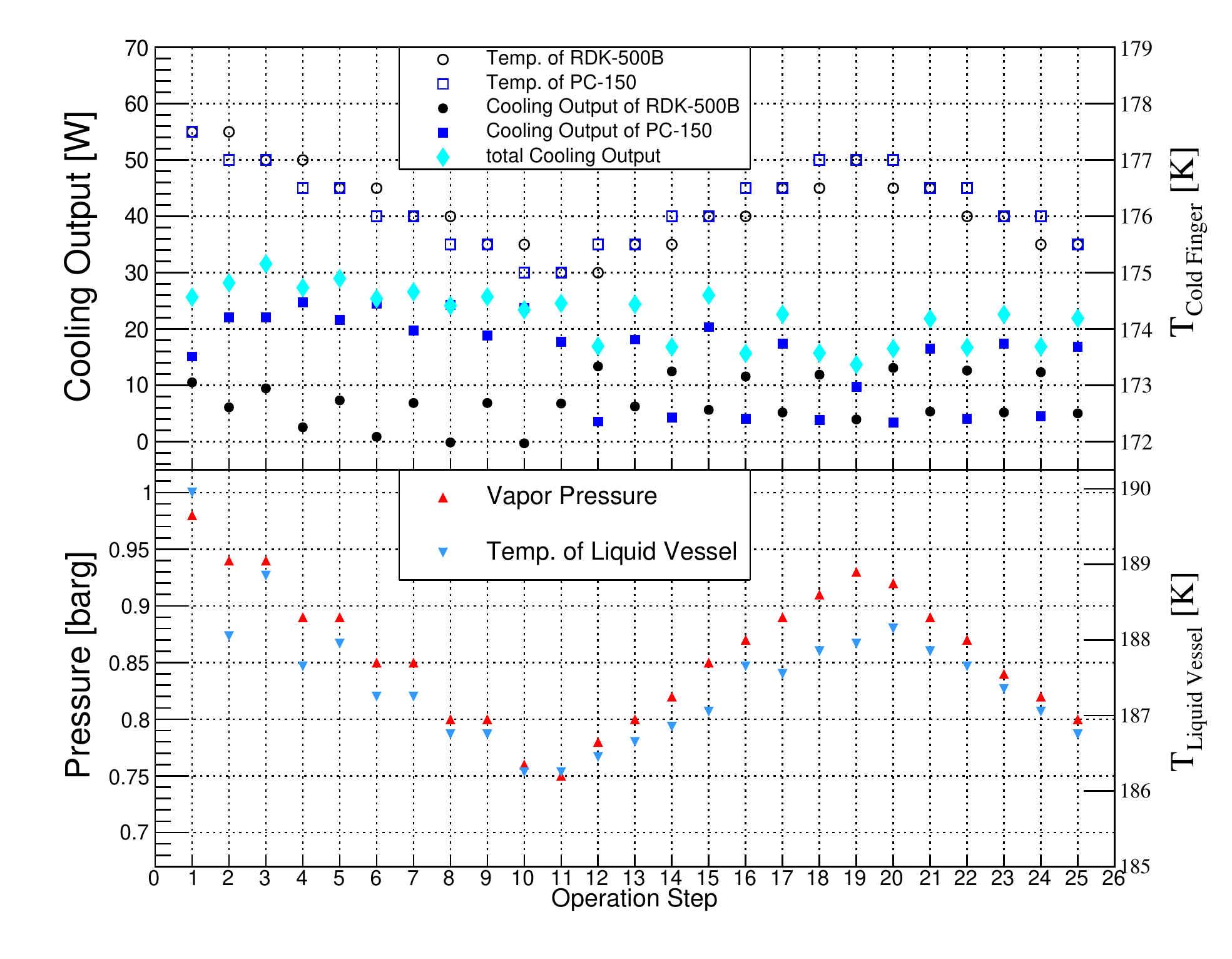}
\caption{The test results of the Cooling Bus with RDK-500B and PC-150 coldheads running.}
\label{fig:working-multi-coldhead}
\end{figure}

The stable condition of the Cryogenic system with one coldhead can be defined simply as the following:
\begin{equation}
  W_{1}=W_{10}+H_{1}+k\times A/L\times\Delta T, \Delta T=T_{i}-T_{1}
\end{equation}
\begin{equation}\label{one coldhead}
  W_{1}=W_{10}+H_{1}+k\times A/L\times(T_{i}-T_{1})
\end{equation}
where $W_{1}$ is the cooling power of the coldhead, $W_{10}$ is passive heat load through the cryostat, $H_{1}$ is the heater power at a given temperature setpoint $T_{1}$, $\Delta T$ is the temperature gradient, $T_{i}$ is the temperature of liquid xenon for the vapor pressure, $k$ is the thermal conductivity of the conductor in the tower, $A$ is the cross section area of the conductor, and $L$ is the length of the conductor. The value of $k$, $A$ and $L$ is constant after the coldhead is fixed on the Cooling Bus. Therefore, the vapor pressure of liquid xenon detector can be adjusted by changing the temperature setpoint $T_{1}$ as $T_{1}<T_{i}$; the effective cooling power($W_{1}-W_{10}$) is measured by the electrical power of the heater.

The condition of two coldheads can be defined as the following:
\begin{equation}
 W_{1}+W_{2}=W_{10}+W_{20}+H_{1}+H_{2}+k\times A/L\times(T_{i}-T_{1})+k\times A/L\times(T_{i}-T_{2})
\end{equation}
\begin{equation}\label{two coldheads}
 W_{12}=W_{120}+H_{12}+2\times k\times A/L\times(T_{i}-\frac{T_{1}+T_{2}}{2})
\end{equation}
where $W_{12}=W_{1}+W_{2}$, $W_{120}=W_{10}+W_{20}$, $H_{12}=H_{1}+H_{2}$. Form of formula~\ref{two coldheads} is the same to that of the formula~\ref{one coldhead} as $\frac{T_{1}+T_{2}}{2}<T_{i}$, the heat exchange area becomes double, therefore, the stable condition with two coldheads should be obtained by setting $T_{12}=\frac{T_{1}+T_{2}}{2}$.

An independent test without the inner vessel was carried out using $\sim$5 kg liquid xenon. The results of the test with two coldheads running simultaneously were shown in Figure~\ref{fig:working-multi-coldhead}. The stabilization time of each operation step was around one hour. The temperature of different copper fingers varied alternatively, so did the cooling output of different coldheads and the vapor pressure.

\begin{figure}[bhtp] \centering
\includegraphics[width=0.8\textwidth]{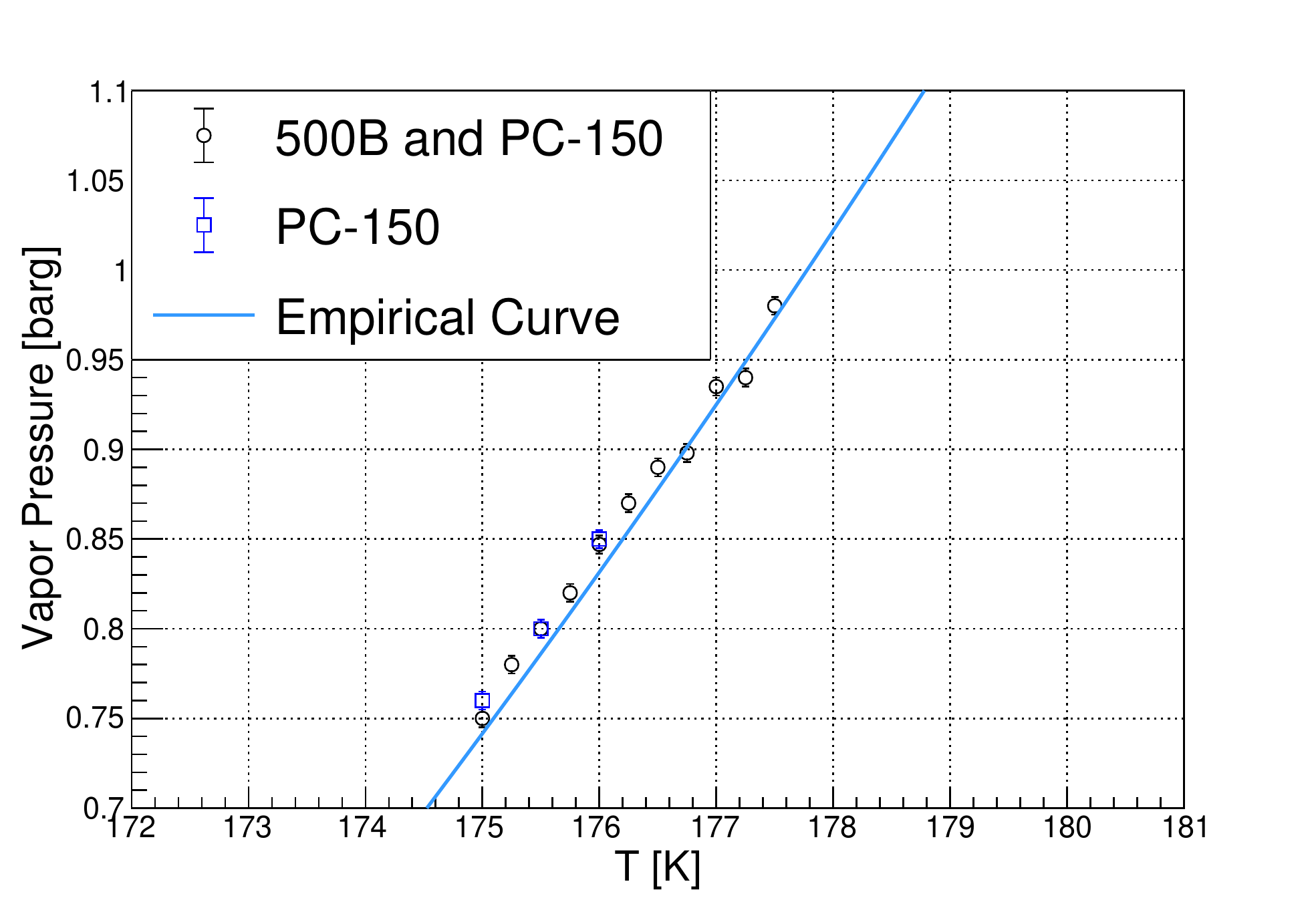}
\caption{The inner pressure VS temperature setpoint T, $T=T_{1}$ for PC-150 as the cooling output of 500B is zero, $T=T_{12}$ for the cooperation of PC-150 and 500B. $T=T_{i}$ for empirical curve of xenon.}
\label{fig:ivp-T}
\end{figure}

Based on the above tests, the following conclusions can be obtained: 1)The vapor pressure can be determined by the temperature setpoints of two coldheads (Figure~\ref{fig:ivp-T}). Moreover, the vapor pressure of two coldheads is consistent with that of one coldhead; 2)The results of operation step 13$\sim$25 show that the cooling output of two coldheads is roughly stable. Their contribution depends on the temperature difference ($T_{i}-T_{1(2)}$); 3)The results of operation step 6, 8 and 10 show that the contribution of 500B coldhead is zero while PC-150 coldhead handles the system alone. It shows that each cooling tower on the Cooling Bus is still relatively independent for liquifying xenon. Due to high thermal resistance of xenon gas, there is no direct heat transfer between the cold fingers.

\begin{figure}[bhtp] \centering
\includegraphics[width=0.8\textwidth]{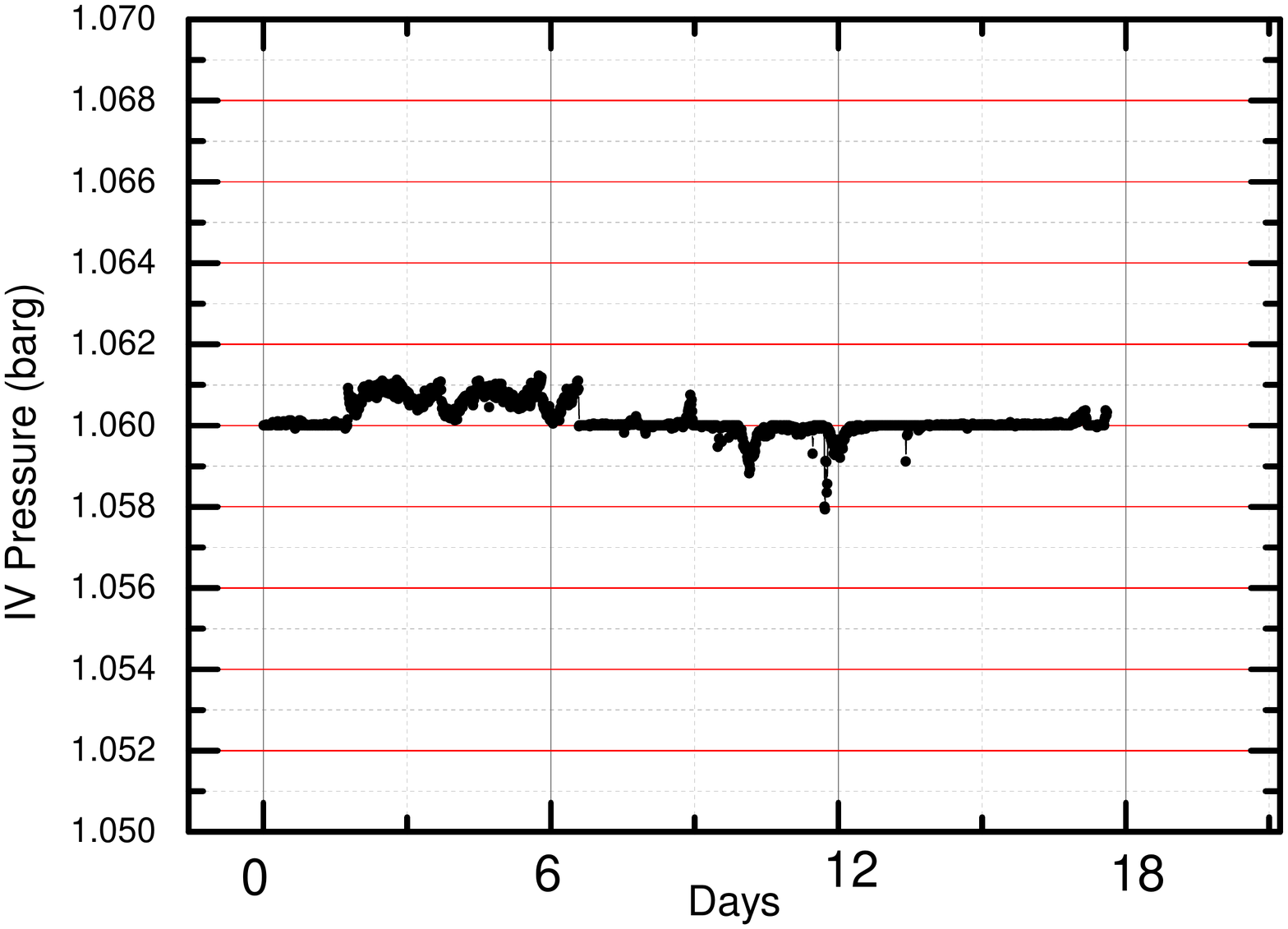}
\caption{The inner pressure vs time as the Cooling Bus with two cooperating coldheads handles about 6 tons of liquid xenon.}
\label{fig:ivp-time-6t}
\end{figure}

The Cooling Bus was then used to stabilize the temperature and pressure of the detector. As shown in Figure~\ref{fig:ivp-time-6t}, the inner pressure was kept stable, and its fluctuation was less than 0.0025~bar. In conclusion, the new designed Cooling Bus with multiple coldheads can work well and different cooling towers can cooperate to handle a large cryogenic detector with huge heat load.

\subsection{Filling and emptying}

\begin{figure}[bhtp] \centering
\includegraphics[width=0.8\textwidth]{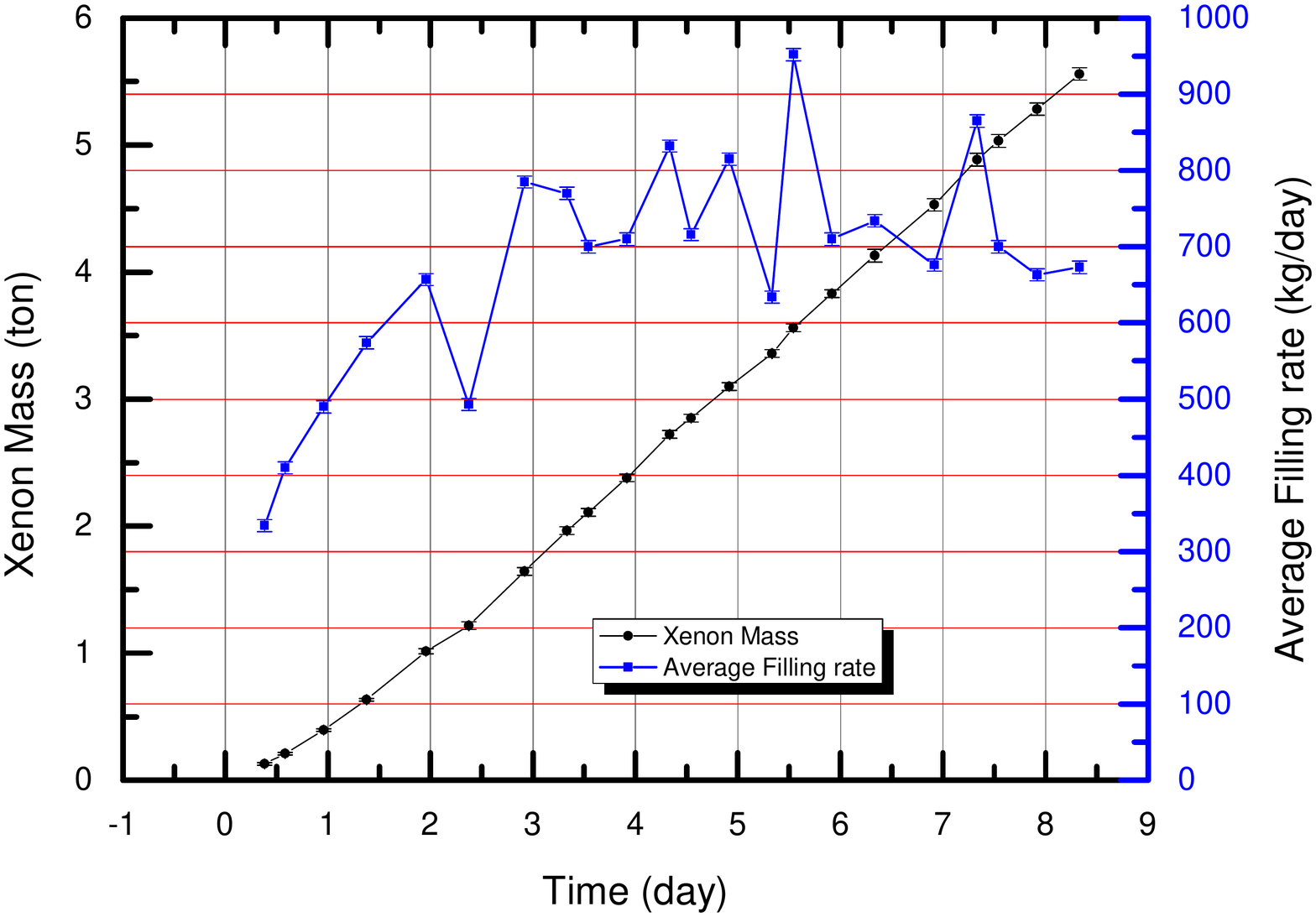}
\caption{Xenon mass vs time for the test filling.}
\label{fig:filling-pandax4t}
\end{figure}

The detector vessels were pumped for a week, then, the inner vessel was filled with 1.5 barg xenon gas. Before starting continuous filling the detector vessel, the inner chamber was pre-cooled with three coldheads for 2 days. The lowest temperature of the inner vessel was around -30\textcelsius~after pre-cooling. Then, the emergency LN2 cooler assisted three coldheads to liquify xenon as fast as possible.

Due to a long distance between xenon storage system and main experimental area, a $\sim $ 30 m gasline(1/2') was used to transfer high pressure xenon gas from the storage to the two parallel gas controller panels of LOOP1 and LOOP2. The filling process with two getters on is shown in Figure~\ref{fig:filling-pandax4t}. The average filling rate is kept around 700 kg/day after 3 liquifying days because of limited heat from the inner vessel. At last, it took about 10 days to liquify $\sim$ 6 tons of xenon.

\begin{figure}[bhtp] \centering
\includegraphics[width=0.8\textwidth]{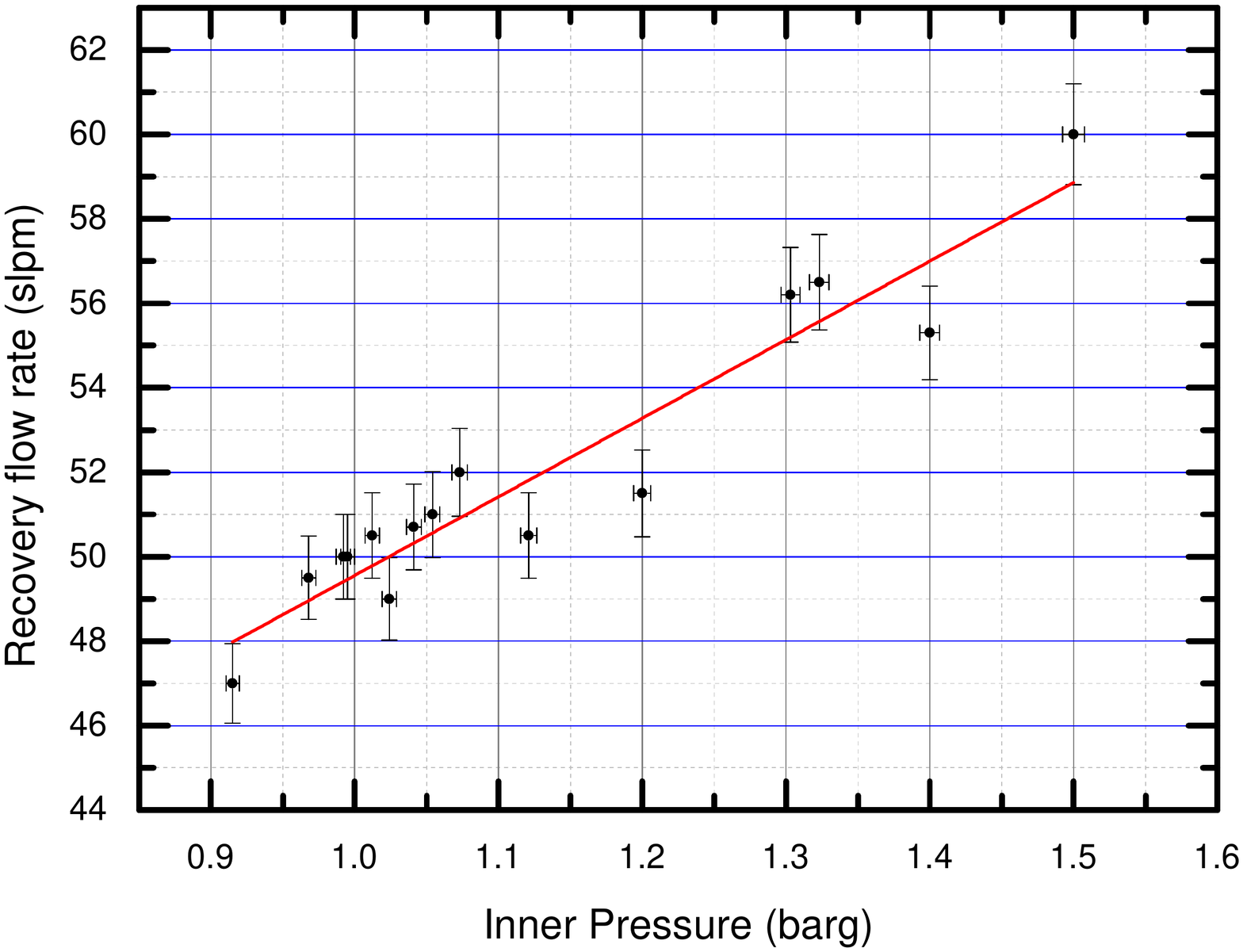}
\caption{The recovery flow rate vs inner pressure, using emergency recovery system.}
\label{fig:recovery-by-LN2-small-iv}
\end{figure}

To empty the vessel at the end of the test, the liquid xenon was
evaporated with alcohol heating pipes, then, the gas xenon went through $\sim$30 m
pipe and was stored by emergency recovery system. Figure~\ref{fig:recovery-by-LN2-small-iv} shows the
recovery flow rate for different inner pressure, higher inner pressure is
helpful to recover xenon faster. The average recovery rate is around 55 slpm
($\sim$ 440 kg/day) for a group of 4 bottles. A high pressure compressor recovery system with faster recovery rate was designed in parallel, which will be reported elsewhere.

\subsection{Heat exchange efficiency of loops}
The maximum liquid xenon
recirculation flow rate of LOOP1 was measured to be $\sim 100$ slpm as the output pressure of KNF pump was about 1.78 barg. That of LOOP2 was $\sim55$ slpm when the output pressure of KNF pump was about 2.53 barg. It shows that its main source of dynamical resistance to the flow
for LOOP2 is the Simpure getter. Therefore, the maximum total flow rate of
LOOP1 and LOOP2 is $\sim$ 155 slpm.

\begin{figure}[bhtp] \centering
\includegraphics[width=0.8\textwidth]{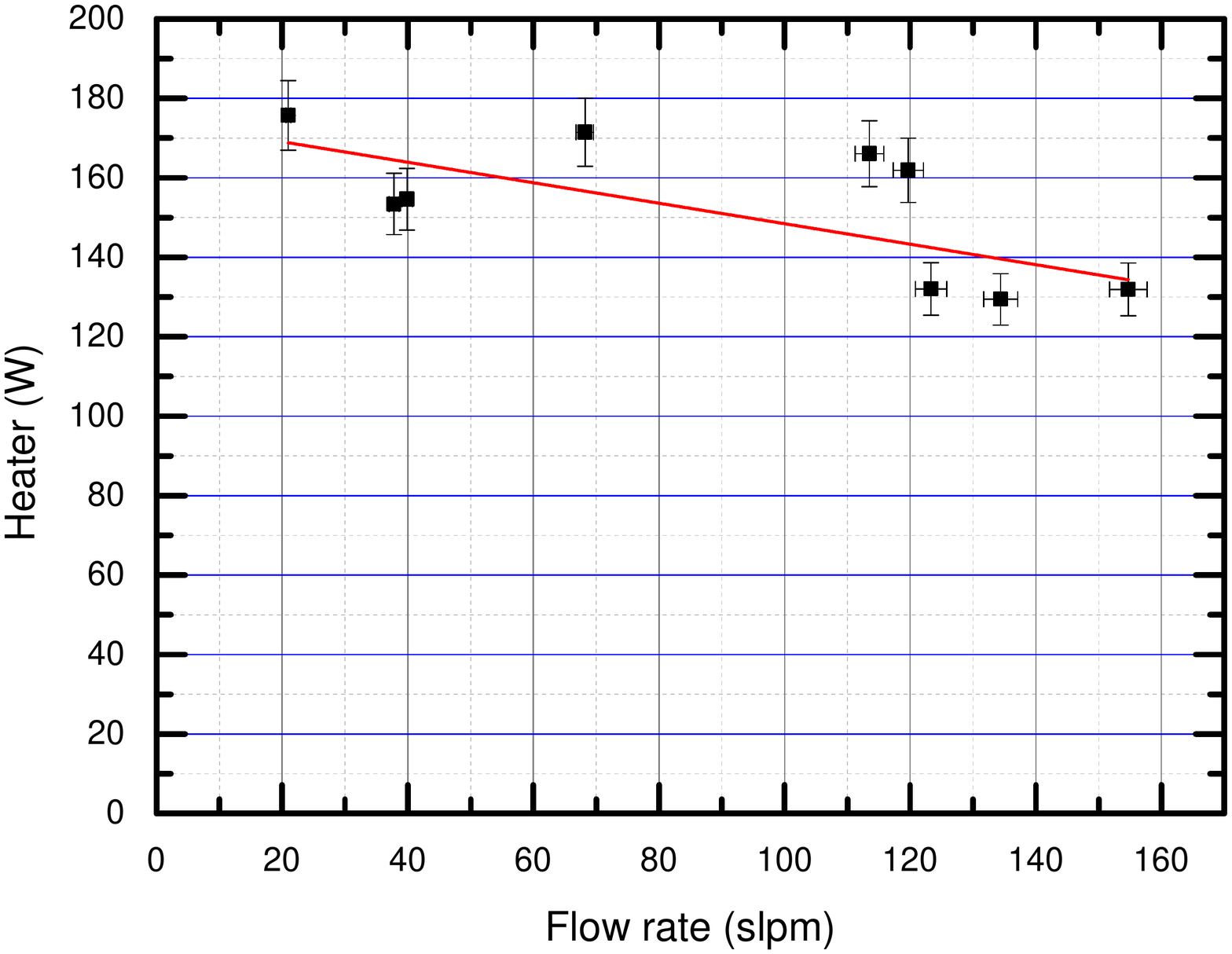}
\caption{The total heating power of two coldheads as a function of total flow rate of LOOP1 and LOOP2 as online purifying.}
\label{fig:ht-flow-pandax4t}
\end{figure}

Figure~\ref{fig:ht-flow-pandax4t} shows the total heater power of two coldheads as a function of the total flow rate of LOOP1 and LOOP2, up
to a rate of 155 slpm, the slope can be interpreted as the inefficiency of
the heat exchanging process. For 6 tons liquid xenon, every operation point is a new equilibrium established, which is time consuming. Here, the fluctuation of these points is larger because of short balance time and difference of flow speeds of the two loops.
The efficiency of the HE can be defined simply as the following\cite{Cry-XENON1T}:
\begin{equation}\label{efficiecny of HE}
  \varepsilon(r)=1-\frac{P(0)-P(r)}{10.74\times r}
\end{equation}
\begin{equation}\label{average efficiecny of HE}
or~~  \overline{\varepsilon}=1+\frac{line~slope(P(r))}{10.74}
\end{equation}
where $r$ is the flow rate (slpm), P(r) is the heater power(W) at a given
circulation flow rate $r$. The value 10.74 W/slpm corresponds to an isobaric
vaporization of liquid xenon at 2 bar followed by an increase in temperature from 178
K to 293 K, of which 8.88 W/slpm is spent on the enthalpy change during
the evaporation\cite{Xenon-Prop}. Here, extra heat into the HE can usually be neglected due to the thick
super insulation. The average efficiency of HEs is $\sim 97.5\pm0.5\%$.
The heat transfer area and flow capability of the HE for PandaX-4T experiment is about 5 $m^{2}$ and 100 $\sim$ 300 slpm respectively (section:\ref{sec:online}). However, the average circulation flow rate of each loop is less than or close to one-third flow capability of HE. Therefore, higher heat exchange efficiency in our new designed loops has been achieved.

\section{Conclusion}
The commission of cryogenics and gas handling for PandaX-4T experiment has been carried out to study and demonstrate the ability of cryogenics at high purification speed. The total cooling power of the Cooling Bus with three cooperating coldheads is $\sim$ 580 W at 178 K. The Cooling Bus with at least two coldheads running can handle $\sim$ 6 tons of liquid xenon even when two loops are running.

The filling rate with the LN2 cooling coil reaches $\sim$ 700 kg/day in the process of continuous liquification of about 6 tons of xenon. For LN2 recovery system, the recuperation rate is $\sim$ 440 kg/day. It will be improved by using a high pressure compressor in the future.

With the use of a high capacity pump and low flow resistance pipes for purification system, the flow rate of LOOP1 reaches $\sim$ 100 slpm and the getter of LOOP2 is the main restriction for the gas flow. The bigger parallel plate HE in the new designed loops provides $\sim 97.5\pm0.5\%$ efficiency of heat exchange. The sum of purification speed is up to $\sim$ 155 slpm for LOOP1 and LOOP2. The LOOP2 could be improved in the future by replacing the Simpure getter with a low flow resistance purifier.

\acknowledgments

This project is supported in part by the Double
First Class Plan of the Shanghai Jiao Tong University,
grants from National Science Foundation of China (Nos.
11435008, 11455001, 11525522, 11775141 and 11755001),
We thank the office of Science and Technology, Shang-
hai Municipal Government (No. 11DZ2260700, No.16DZ2260200, No. 18JC1410200) and the Key Laboratory for Particle Physics, Astrophysics and Cosmology, Ministry of Education, for important support. Finally, we thank the CJPL administration and the Yalong River Hydropower Development
Company Ltd. for indispensable logistical support and other help.


\end{document}